\documentclass{article}
\usepackage{ijcai11}

\usepackage{times,graphicx}

\usepackage{array}
\newcolumntype{M}[1]{>{\centering\arraybackslash}m{#1}}

\title{Modeling opinion leader's role in the diffusion of innovation\\\strut\\\small{\emph{Internship report originally written in June 2018 by intern N. Vodopivec under the supervision of C. Adam and J.-P. Chanteau}}}
\author{Nata\v{s}a Vodopivec \\
\small{Univ Grenoble-Alpes}\\ \small{Grenoble INP} \And Carole Adam \\
\small{Univ Grenoble-Alpes}\\\small{Grenoble Informatics Lab} \And Jean-Pierre Chanteau \\ \small{Univ Grenoble-Alpes}\\ \small{Centre de Recherche en Economie de Grenoble}}

\begin{document}

\maketitle

\renewcommand{\arraystretch}{1.4}

\begin{abstract}
The diffusion of innovations is an important topic for the consumer markets. Early research focused on how innovations spread on the level of the whole society. To get closer to the real world scenarios agent based models (ABM) started focusing on individual-level agents. In our work we will translate an existing ABM that investigates the role of opinion leaders in the process of diffusion of innovations to a new, more expressive platform designed for agent based modeling. We will do it to show that taking advantage of new features of the chosen platform should be encouraged when making models in the field of social sciences in the future, because it can be beneficial for the explanatory power of simulation results.
\end{abstract}

\section{Introduction}
Diffusion refers to the process by which an innovation is adopted over time by members of a social system. An innovation commonly refers to a new technology, but it can be understood more broadly as a spread of ideas and practices \cite{kiesling}. The question whether a certain innovation will diffuse in society successfully or not has always been of important nature at the market level and has gained interest of many researchers since a number of pioneering works appeared in the 1960s.
\paragraph{} From the marketing perspective, it is of great importance to understand how information starting from mass media and traveling through word-of-mouth WoM affects adoption decisions of customers and consequently the diffusion of a new product \cite{vanEck}. Mass media takes the role of an external influence to a society and WoM the role of an internal influence within the society. Traditionally, models were based on \textit{macro level} looking at the society as a whole.

Most such aggregate models stem from the model introduced by Bass~\shortcite{bass}, which takes the structure of a basic epidemic model where diffusion of innovation is seen as a contagious process driven by external and internal influences. This model assumes that the market is homogeneous, which means that all customers have the same characteristics. It further assumes that each consumer is connected with all other consumers and can thus influence all others. From these two assumptions it follows that the probability of adopting is linearly related to the number of past adopters.
These assumptions are limitations of aggregate models as they ignore that in the real world consumers are individuals and as such heterogeneous and a part of complex social structures.

\paragraph{}
To try to overcome these limitations, agent-based modeling \cite{macal} has been increasingly adopted in the recent times. Agent based modeling takes a different approach to diffusion of innovations, because it looks at the society from the \textit{micro level}. Here, an observed entity is not a society as a whole, but an individual, represented as an agent. Customers' heterogeneity, their social interactions and their decision making process can be modeled explicitly \cite{kiesling}. When simulated, macro level observations of network changes emerge from the micro level interactions between the individuals.

Most agent based models of innovation diffusion have a similar structure and comprise of the following elements \cite{jensen}:
\begin{enumerate}
\item \textit{Consumer agents} define the individual entities that can adopt an innovation. These can be individual persons, households, or groups of households. They are heterogeneous.
\item \textit{Social structure} is a description of connections between singular agents, dividing them in different consumer groups.
\item \textit{Decision making processes} are the key actions of consumer agents in any social model, by which agents decide to adopt or reject the innovation.
\item \textit{Social influence} between agents often affects decision making processes and is commonly modeled as a social network graph. Models vary in the range at which social influence is exceeded. This can be influence from direct peers, from the respective social group or the entire population of agents. All these ranges of influence can be modeled as a social network graph.
\end{enumerate}

\begin{figure*}
    \centering
    \includegraphics[scale=0.4]{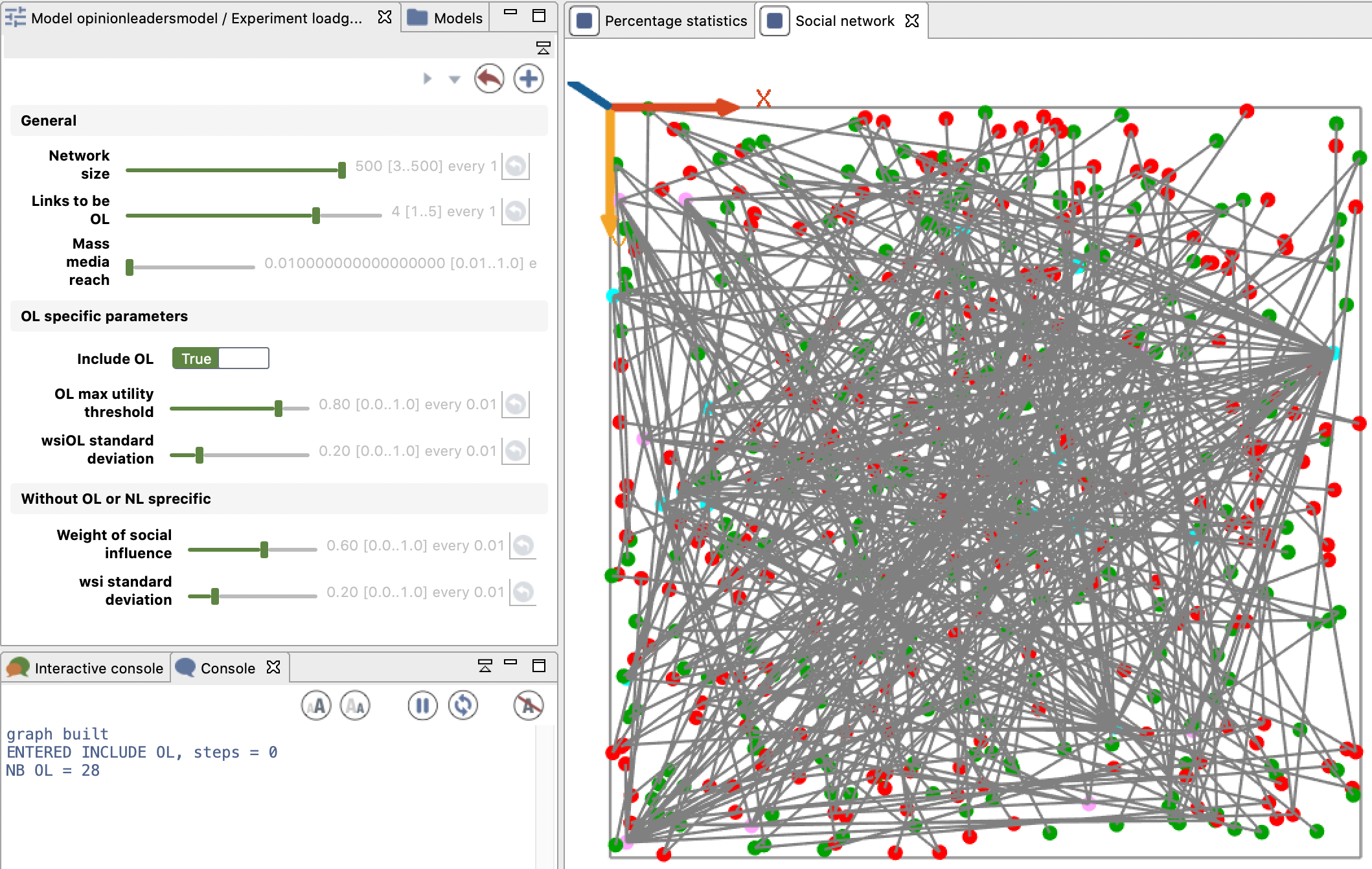}
    \caption{Screenshot of the GAMA simulator}
    \label{fig:gama}
\end{figure*}

\paragraph{}
We are interested in modeling and simulating how innovation, both in the sense of the ideas, behaviours and in the sense of the products, spreads in the population. We have chosen to implement the model on a GAMA platform, which is seen as a current state-of-the-art agent-based modeling language and as an improved successor of the NetLogo platform. Figure~\ref{fig:gama} presents a screenshot of the implemented simulator, showing the social network with opinion leaders (in pink) and adopters of the innovation (in green).

The main challenge when modeling social models is the verification of the final model. The verification can be done either using strong theoretical support or using obtained empirical data. However, due to the scope of our project it was difficult to obtain either. This is the reason why we have chosen to use an existing NetLogo model to rewrite and improve in GAMA, because this way we would be able to validate our model against it. A study by van Eck~\shortcite{vanEck} (further use: reference study) was picked as it not only models the diffusion of innovations, but additionally investigates the role of opinion leaders in the process, which is another interesting phenomenon. Aside from agents being heterogeneous, they are further divided into two groups, namely the influentials or \textit{opinion leaders (OL)} and followers or \textit{non-leaders (NL)}.

\paragraph{}
Goldenberg et al~\shortcite{goldenberg} determine influentials by three factors: connectivity, knowledge and personality characteristics. Opinion leaders are a type of influential customers that have all of the characteristics of the influentials represented as central positions in the network (which means high connectivity), market knowledge (not necessarily about a specific product but about markets in general) and innovative behaviour.

The reference study uses four critical assumptions about opinion leaders, which are later successfully checked by an empirical study: (1) OL have more contacts, (2) OL possess different characteristics, (3) OL exert different types of influence and (4) OL are among earlier adopters.

Two important characteristics of opinion leaders are their innovativeness and their interpersonal influence. Regarding the degree of innovativeness, it means that opinion leaders have more experience and expertise with the product category than the other consumers and that they have been exposed to more information \cite{lyons}.
Two main types of interpersonal influence exist:
\begin{itemize}
\item \textit{Informational influence} is the tendency to accept information from others and believe it. Opinion leaders influence other consumers by giving them advice about a product.
\item \textit{Normative influence} stems from the people's tendency to follow a certain norm; to adopt a product in order to be approved by other consumers. Normative influence can also be referred to as a social pressure.
\end{itemize}

Reference study assumes that opinion leaders play an important role in both diffusion of information about products (informational influence) and the diffusion of products themselves (i.e. more product adoptions result in normative influence) \cite{vanEck}. Therefore the influence of the opinion leaders on the speed of diffusion of both information and product, and on maximum adoption percentage in the process of diffusion of innovation, is investigated.

\paragraph{}
The focus of this study is to investigate the speed of information diffusion, the speed of product diffusion, and the maximum adoption percentage of the product.

The article is structured as follows: in Section \ref{sec:hypoth} we describe the hypotheses that the reference study has set up and verified; Section \ref{sec:model} introduces the model with its agents, parameters, and social network; in Section \ref{sec:results} we present the experiments settings and discuss our simulation results; and in Section \ref{sec:conclusions} we address the conclusions and suggestions for further work.

\section{Hypotheses}
\label{sec:hypoth}

While investigating the role of opinion leaders in the innovation diffusion process, the impact of each of its three characteristics (innovative behaviour, normative influence, market knowledge) is looked at and thus more hypotheses are set up. We have chosen to validate our model against the following hypotheses put forward and successfully proven in the reference study \cite{vanEck}:
\subparagraph{H1a:} "The more innovative behaviour of the opinion leader results in a higher adoption percentage."
\subparagraph{H1b:} "If the weight of normative influence becomes more important to followers, the increase in the adoption percentage caused by the more innovative behavior of opinion leaders increases."
\subparagraph{H2a:} "Opinion leaders are less sensitive to normative influence than are followers."
\subparagraph{H2b:} "If opinion leaders are less sensitive to normative influence, adoption percentages increase."
\subparagraph{H2c:} The less sensitive opinion leaders are to normative influence, the more the adoption percentages increase.
\subparagraph{H3a:} "Opinion leaders are better at judging product quality, which results in a higher speed of information diffusion."
\subparagraph{H3b:} "Opinion leaders are better at judging product quality, which results in a higher speed of product diffusion."

\section{Model}
\label{sec:model}
In this chapter we describe in more detail how our model was built.

\subsection{Network}

Bohlman et al.~\shortcite{bohlman} indicate that specific network topologies in agent based modeling strongly influence the process of innovation diffusion: they affect the likelihood that diffusion spreads and the speed of adoption. This is because the network topology specifies the location and the number of links of innovators. A scale-free network structure proposed by Barabasi and Albert (1999) is used, because it stems from many empirical researches and is confirmed to imitate real world societies where some agents serve as hubs, meaning their number of connections greatly exceeds the average and they have central positions in the network.

\subsection{Agents}

Each agent is described by the following attributes:
\begin{itemize}
\item \textit{Opinion leader} tells whether an agent is an opinion leader or a non-leader.
\item \textit{Quality threshold} is a value given randomly uniformly to each agent before the beginning of a simulation. Its values are uniformly distributed in the range U(0,1).
\item \textit{Known quality} describes what the agent currently thinks of the product quality. This value is set dynamically when the agent gets aware of the product or adopts it, as is further explained in Section \ref{subsec:experiment}.
\item \textit{Utility threshold} is a value given randomly uniformly to each agent before the beginning of a simulation. OL and NL have different ranges from where this value can be taken. For NL it is U(0,1) and for OL it is U(0, \textit{max}), where the maximum value is defined by a parameter of the experiment.
\item \textit{Awareness} tells whether an agent is aware of the product or not.
\item \textit{Adopted} tells whether an agent has adopted the product or not.
\item \textit{Weight of normative influence} is different for OL and NL and values are set dynamically as a normal distribution where average value and standard deviation are set as parameters of the simulation.
\end{itemize}

The uniform distribution of the values of utility thresholds and of quality thresholds for individual agents makes the population heterogeneous.

\paragraph{}
An agent's decision to adopt is based on its utility threshold. The agent's utility is calculated at each iteration of the simulation, and once it passes the agent's utility threshold the agent adopts the innovation. The utility function consists of a weighted sum of the \textit{individual preference} and the \textit{social influence}. First represents informational influence and describes the agent's opinion on the product quality, and second represents normative influence and takes into account the number of neighbouring agents that have already adopted the product. When the weight of the social influence of a certain agent is low the agent is very individualistic and is consequently hardly influenced by neighbours. On the contrary, high weight value means that the agent is very socially susceptible \cite{vanEck}.

\subsection{Parameters}

The model contains several parameters, which describe the influence of opinion leaders in various market settings \cite{vanEck}. Some parameters are fixed for all experiments and others vary experimentally. The group of fixed parameters and their values, derived from the model by Delre et al.~\shortcite{delre2}, are presented in Table \ref{tab:fixed}. The product quality is set to 0.5, meaning that if the agents base their decisions to adopt a product purely on their individual preferences, approximately 50\% will never adopt. The mass media coefficient was set from prior studies. It represents a strong mass media support because many of the agents in the empirical study were reached by mass media (i.e. one percent of population is reached in each step).

\begin{table}
	\begin{center}
		\begin{tabular}{p{3.5cm} M{1.8cm} M{1.8cm}}
			\hline
			Variable & Parameter & Value \\
			\hline
			Product quality & \textit{q} & 0.5 \\
            Mass media coefficient & \textit{m\_m} & 0.01 \\
			Number of agents & \textit{nb\_agents} & 500 \\
			\hline
		\end{tabular}
        \caption{Settings for global parameters, that were fixed in current experiments}
        \label{tab:fixed}
	\end{center}
\end{table}

\begin{table}
	\begin{center}
		\begin{tabular}{p{4.5cm} M{1.3cm} M{1.3cm}}
			\hline
			Variable & Parameter & Value  \\
			\hline
			Max utility threshold of OL & \textit{max} & 0.8 \\
            Average normative \newline influence, OL & \textit{avg\_ni\_ol} & 0.51 \\
            Standard deviation for normative influence, OL & \textit{dev\_ni\_ol} & 0.2 \\
            Average normative \newline influence, NL & \textit{avg\_ni\_nl} & 0.6 \\
            Standard deviation for normative influence, NL & \textit{dev\_ni\_nl} & 0.2 \\
            OL judges product better & NA & Yes \\
			\hline
		\end{tabular}
        \caption{Settings for base model parameters}
        \label{tab:params_varied}
	\end{center}
\end{table}

The varied parameters are changed one at a time per experiment to test the separate hypotheses. The parameters and their values, derived from empirical study conducted by reference study are presented in Table \ref{tab:params_varied}. First a base experiment with these values was run so that later hypotheses could be tested realistically. The innovativeness of opinion leaders is implemented as smaller possible values of it's utility threshold with regard to that of the followers (the utility threshold of the followers has a uniform distribution in the range U(0, 1.0), for OL it's in the range U(0, 0.8)), which makes them approximately 20\% more likely to adopt the product. The difference is not big as OL are trying to avoid being too innovative, because if they adopted a product that turned out to be unsuccessful, they would loose followers. As observed in the empirical study, the weight of normative influence of opinion leaders holds a lower value ($\beta_{OL}=$ 0.51) than that of the followers ($\beta_{NL} =$ 0.6) as they care less about the social pressure. The weights of normative and informative influences sum up to 1, so the weight of informative influence is 1 - $\beta$.
The model can be run either with opinion leaders in the network or without them. This was important to be able to see whether the diffusion of the innovation indeed spreads faster in the networks where innovative opinion leaders are present.

\begin{table*}
	\begin{center}
		\begin{tabular}{p{4.5cm} M{2.5cm} M{2.5cm} M{2.5cm} M{2.5cm}}
			\hline
			 & Innovativeness of OL & Weight of normative influence OL & Weight of normative influence NL & \\
            \cline{2-4}
            Model (hypothesis tested with model) & $U_{i, min}$ & $\beta_{i, OL}$ & $\beta_{i, NL}$ & Quality of the product judgment (OL) \\
			\hline
            \textit{Base Model 1} & \textbf{U(0, 0.8)} & \textbf{N(0.51, 0.2)} & \textbf{N(0.6, 0.2)} & \textbf{Yes} \\
            Model 2 (H1a) & \textbf{U(0, 1)} & N(0.51, 0.2) & N(0.6, 0.2) & Yes \\
            Model 3 (H1b) & U(0, 0.8) & N(0.51, 0.2) & \textbf{N(0.8, 0.2)} & Yes \\
            NA (H2a) & NA & NA & NA & NA \\
            Model 4 (H2b) & U(0, 0.8) & \textbf{N(0.57, 0.2)} & \textbf{N(0.57, 0.2)} & Yes \\
            Model 5 (H2c) & U(0, 0.8) & \textbf{N(0.2, 0.2)} & N(0.6, 0.2) & Yes \\
            Model 6 (H3a, H3b) & U(0, 0.8) & N(0.51, 0.2) & N(0.6, 0.2) & \textbf{No} \\
            \hline
		\end{tabular}
        \caption{Parameter settings for hypotheses (adapted from \protect\cite{vanEck})}
        \label{tab:params_hypoth}
	\end{center}
\end{table*}

\section{Experiments and results}
\label{sec:results}
In this chapter we first present the experiment settings, then discuss the results and finally do a comparison between NetLogo and GAMA platforms.

\subsection{Experiment settings}
\label{subsec:experiment}

A model was created for each separate hypothesis. The values of the varied parameters used for each model are shown in Table \ref{tab:params_hypoth}. Each model was run in a separate experiment that consisted of 25 time steps, which was enough for the maximum adoption percentage to be reached. To collect results for statistics each experiment was run with the same settings 60 times. We realize that 60 is a low number of repetitions for completely adequate statistics, but we faced a problem of the GAMA platform freezing due to too big memory consumption while trying to run it in batch mode, where more than one experiment is run one after another automatically. We did not anticipate this to happen as the calculations were very fast when running 500 consecutive experiments in NetLogo and GAMA is seen as it's improved successor. Thus, we were reduced to having to run each experiment manually which proved to be quite time consuming so we limited the number of runs to 60 and might do more tests to calibrate the results if needed in the future.

Each time step further consisted of three phases: mass media, WoM and adoption. In the beginning of the experiment no agents are aware of the product or have adopted it. Then mass media informs a predefined percentage (in our case 1\%) of the population about it. In this step, the better market knowledge of the opinion leaders is implemented as such: because opinion leaders are able to make good product judgment they will have learned of a real product quality from mass media and their quality judgment will become equal to it (\textit{q} = 0.5, Table \ref{tab:fixed}). On contrary, followers are not able to make this judgment so they become aware of the product but their perceived product quality gets a random value. The followers are able to learn about the real product quality only by WoM from trusted sources, that is from opinion leaders and agents who have already adopted the product.
In the word of mouth stage, agents talk with their neighbors and may learn about the real product quality if their neighbors are certain about it.
In the adoption stage, agents can decide to adopt the product if they are aware about it and if the current value of utility function exceeds their utility threshold.

\subsection{Results and Validation}

\begin{table*}
	\begin{center}
        \begin{tabular}{lcccccc}
            \hline
			& \multicolumn{2}{c}{\begin{tabular}[c]{@{}c@{}}Adoption percentage\\ (standard deviation)\end{tabular}} & \multicolumn{2}{c}{\begin{tabular}[c]{@{}c@{}}Speed of information diffusion\\ Average number of steps\\ (standard deviation)\end{tabular}} & \multicolumn{2}{c}{\begin{tabular}[c]{@{}c@{}}Speed of product diffusion\\ Average number of steps\\ (standard 						 deviation)\end{tabular}} \\
			\cline{2-7}
            & \begin{tabular}[c]{@{}c@{}}Reference\\ study\end{tabular} & \begin{tabular}[c]{@{}c@{}}Our\\ study\end{tabular} & \begin{tabular}[c]{@{}c@{}}Reference\\ study\end{tabular} & \begin{tabular}[c]{@{}c@{}}Our\\ study\end{tabular} & \begin{tabular}[c]{@{}c@{}}Reference\\ study\end{tabular} & \begin{tabular}[c]{@{}c@{}}Our\\ study\end{tabular} \\
			\hline
            \textit{Base Model 1 - no OL} & 0.398 (0.05) & 0.401 (0.05) & 3.64 (1.5) & 4.43 (1.47) & 6.27 (2.0) & 5.78 (1.51) \\
            \textit{Base Model 1} & 0.491 (0.05) & 0.454 (0.05) & 1.75 (1.2) & 1.96 (0.43) & 4.94 (1.2) & 2.80 (0.78) \\
            Model 2 (H1a) & 0.405 (0.04) & 0.398 (0.05) & & & & \\
            Model 3 (H1b) & 0.458 (0.06) & 0.395 (0.06) & & & & \\
            NA (H2a) & & & & & & \\
            Model 4 (H2b) & 0.480 (0.05) & 0.455 (0.06) & & & & \\
            Model 5 (H2c) & 0.515 (0.04) & 0.488 (0.05) & & & & \\
            Model 6 (H3a, H3b) & & & 4.73 (2.22) & 4.11 (1.70) & 7.76 (2.21) & 5.64 (1.65) \\
			\hline
		\end{tabular}
        \caption{Results of tests of each hypothesis}
        \label{tab:results}
	\end{center}
\end{table*}

Before looking at the models testing the hypotheses we had to make sure that our model confirms the base assumption. It claims that in networks that include opinion leaders, higher speed of both the information and product diffusion as well as greater adoption percentage are achieved, than in the networks without them. Thus, we ran the base model in two experiments, once with opinion leaders and once without them. The weight of normative influence ($\beta_i$) in the comparison model with no opinion leaders is 0.75 (obtained from reference study). The average values of the results and their standard deviations for these two tests as well as for the rest of the tests can be found in Table \ref{tab:results}. We can see that in the model with opinion leaders the information diffuses faster than in the model without the opinion leaders, in the first it takes 1.96 steps compared with 4.43 in the other. The same case happens for the speed of product diffusion, in the model with OL it takes 2.80 steps compared to 5.78 steps in the model without the OL, which means that the product diffuses faster in the model with OL. Thirdly, the value of the average adoption percentage is higher in the model with OL (0.45 in the model with OL and 0.40 in the model without), which also confirms our assumptions. Therefore, the base model successfully proves that the opinion leaders product higher speeds of information and product diffusion and higher adoption percentage.

Table \ref{tab:results} shows the obtained averaged results from the reference study and from our model for each of the hypotheses models. Each hypothesis was run in a different experiment on it's own model, for which the values are presented in the Table \ref{tab:params_hypoth}, except for the hypothesis H2a which the reference study validated by empirical study. When comparing the results we can see that even though values are a bit different, their proportions stay the same, meaning that our model was successfully validated against the reference NetLogo model and as such that the same as in the NetLogo model the hypotheses Ha1, H2a (empirical study), H2b, H2c, H3a and H3b got supported while the hypothesis H1b did not get supported.

\subsection{Comparing the platforms}
The differences might be partially attributed to the smaller sample sizes that we use to average the results, but we think they're mostly the reason of a different execution flow in the GAMA platform. It is here that GAMA platform introduces a difference that we find important when making social models. The execution flow of NetLogo for the model of diffusion of innovation is sequential and iterative. For each of the 25 steps of the simulation the three stages (mass-media, word of mouth, adoption) are executed one after another, where first one has to complete for all of the agents before the next stage can commence. Inside each stage, the agents execute the actions linked to it  iteratively in a loop, the agent 1 does it first and agent 500 the last. The agents inherently act  as small blocks of non-connected code and the order of their execution can never be different as the loop over the agents that calls each of them determines it.

On the other hand, on GAMA platform each agent acts as it's own entity with it's own behaviours. During the simulation of the 25 steps, the only role of the world agent that stands above all other agents (on GAMA platform the world agent acts similar than a main function in many programming languages) is to schedule them, i.e. gives them an opportunity to act. While the agents still do not all run at the same time in parallel, they are not connected with actions of the other agents. When each agent gets it's turn it runs it's behaviours, which are in turn mass-media, WoM and adoption. So the prime difference is that on GAMA platform the main program only calls the agents and after that is has no control over how they execute, they act as individual entities. 

However, in our model the world agent mostly still calls the consumer agents iteratively, starting with agent 1 and finishing with agent 500, which is still not very representative of the real world because the order of the agents is the same at each simulation step. There exists a solution to this problem which is discussed in section \ref{sec:conclusions}. 

\section{Conclusions and Further research}
\label{sec:conclusions}

We have successfully established and validated a model of diffusion of innovation nn the state-of-the-art GAMA platform that is designed for agent-based modeling. We think that it is an important step to take towards the more realistic modeling of social interactions. However, as mentioned before in Section \ref{sec:results} the agents still get executed in the same order in each step of the simulation. This could lead to unrealistic simulations. We would like to highlight one example of this problem, namely discuss the execution of the Word of Mouth stage. When in this stage, the agent talks to all of its neighbours, and if any of them know of the true product quality, then the agent becomes aware of it by WoM. Now imagine the first simulation step when after the mass-media stage at most 1\% of the population has become aware of the product. During the WoM stage agents will be called upon iteratively and each of them will have larger probability that some of its previously non-aware neighbours have now become aware and can thus share their knowledge about the product. Consequently, in each of the 25 steps of the simulation, agent number 1 will always have lesser probability to become aware by WoM than agent number 500. 

To solve this issue GAMA platform provides an option of calling agents in a random (\textit{shuffled}) order. In NetLogo such option could be implemented manually, but would be hard to achieve. As a future work we think that adding this property and observing the obtained results could be a good idea. The results might stay the same, but the micro structure of the model would become closer to the real world social models.

Another promising option of further research would be the extension of the current agents to BDI agents. Agent based modeling is already a step forward from the old aggregate models where humans were modeled as equal homogeneous entities. However, when handling ABMs in the field of social sciences, human agents can be further improved to become more human-like by giving them personality traits. These agents are called belief, desire and intention (BDI) agents. The model allows to use more complex and descriptive agent models to represent humans. It attempts to capture common understanding of how humans reason through: \textit{beliefs} which represent the individual's knowledge about the environment and about their own internal state; \textit{desires} or more specifically \textit{goals} (non-conflicting desires which the individual has decided they want to achieve); and \textit{intentions} which are the set of plans or sequence of actions which the individual intends to follow in order to achieve their goals \cite{adam}. Two other important functionalities a BDI system must have are a \textit{rational process} by which an agent decides which intentions to follow depending on the current circumstances, and the level of \textit{commitment} to the set of intentions to achieve a long-term goal.

We think that BDI agents are important to give higher descriptive value on results of social studies, which a diffusion of innovation certainly is. They give more information on how agents behave and a deeper insight on how innovation diffuses in the population. As a future work, we will upgrade this model by expanding its agents to BDI agents, now that the model has been translated to GAMA, which allows BDI architecture. We will then add different human factors to these agents and observe how they affect the spread of the diffusion of an innovation and it's speed and whether the results will stay in line with the original model.

\appendix
\nocite{delre}

\bibliographystyle{named}

\bibliography{ijcai11}

\begin{thebibliography}{}

\bibitem[\protect\citeauthoryear{Adam and Gaudou}{2016}]{adam}
Carole Adam and Benoit Gaudou.
\newblock BDI agents in social simulations: a survey.
\newblock {\em The Knowledge Engineering Review}, 31(3):207--238, Cambridge University Press, 2016.

\bibitem[\protect\citeauthoryear{Bass}{1969}]{bass}
F. M Bass.
\newblock A new product growth for model consumer durables.
\newblock {\em Management Science}, 15(5):215--227.

\bibitem[\protect\citeauthoryear{Bohlman \bgroup \em et al.\egroup}{2010}]{bohlman}
J. Bohlman, R. Calantone and M. Zhao.
\newblock The effects of market network heterogeneity on innovation diffusion: An agent-based modeling approach.
\newblock {\em Journal of Product Innovation Management}, 27(5):741--60.

\bibitem[\protect\citeauthoryear{Delre \bgroup \em et al.\egroup
  }{2010}]{delre}
Sebastiano A. Delre.
\newblock Will it spread or not?: The effects of social influences and network topology on innovation diffusion.
\newblock {\em The journal of product innovation management : an international publication of the Product Development \& Management Association}, 27(2), 2010.

\bibitem[\protect\citeauthoryear{Delre \bgroup \em et al.\egroup
  }{2007}]{delre2}
Sebastiano A. Delre, W. Jager, T. H. A Bijmolt and M. A. Janssen.
\newblock Targeting and timing promotional activities: An agent-based model for the takeoff of new products.
\newblock {\em Journal of Business Research}, 60(8):826--35.

\bibitem[\protect\citeauthoryear{van Eck \bgroup \em et al.\egroup
  }{2011}]{vanEck}
Peter S. van Eck, Wander Jager and Peter S. H. Leeflang.
\newblock Opinion leaders' role in innovation diffusion : a simulation study.
\newblock {\em The journal of product innovation management : an international publication of the Product Development \& Management Association.}, 28.2011(2):187--203, Oxford, Blackwell Publishing, 2011.

\bibitem[\protect\citeauthoryear{Goldenberg \bgroup \em et al.\egroup
  }{2009}]{goldenberg}
J. Goldenberg, S. Han, D. R. Lehmann and J. W. Wong.
\newblock The role of hubs in the adoption process.
\newblock {\em Journal of Marketing}, 73(2):1--13.

\bibitem[\protect\citeauthoryear{Jensen and Chappin}{2017}]{jensen}
Thorben Jensen and Emile J.L. Chappin.
\newblock Automating agent-based modeling: Data-driven generation and application of innovation diffusion models.
\newblock {\em Environmental Modelling \& Software}, 92:261--268, 2017.

\bibitem[\protect\citeauthoryear{Kiesling \bgroup \em et al.\egroup
  }{2012}]{kiesling}
Elmar Kiesling, Markus Günther, Christian Stummer and Lea M. Wakolbinger.
\newblock Agent-based simulation of innovation diffusion: A review.
\newblock {\em Central European Journal of Operations Research}, 20:183--230, 2012.

\bibitem[\protect\citeauthoryear{Laciana \bgroup \em et al.\egroup
  }{2017}]{laciana}
Carlos E. Laciana, Gustavo Preyra and Santiago L. Rovere.
\newblock Size invariance sector for an agent-based innovation diffusion model.
\newblock {\em ARXIV}, 1706.03859, 2017.

\bibitem[\protect\citeauthoryear{Lyons and Henderson}{2005}]{lyons}
B. Lyons and K. Henderson.
\newblock Opinion leadership in a computer-mediated environment.
\newblock {\em Journal of Consumer Behavior}, 4(5):319--29.

\bibitem[\protect\citeauthoryear{Macal and North}{2005}]{macal}
C. M. Macal and M. J. North.
\newblock Tutorial on agent-based modeling and simulation.
\newblock {\em In 37th Winter Simulation Conference. Introductory Tutorials: Agent-Based Modeling}, 2--15.

\bibitem[\protect\citeauthoryear{Mills and Schleich}{2012}]{mills}
Bradford Mills and Joachim Schleich.
\newblock Residential Energy-Efficient Technology Adoption, Energy Conservation, Knowledge, and Attitudes: An Analysis of European Countries.
\newblock {\em Energy Policy}, 49, 2012.

\bibitem[\protect\citeauthoryear{Zhang and Vorobeychik}{2016}]{zhang}
Haifeng Zhang and Yevgeniy Vorobeychik.
\newblock Empirically Grounded Agent-Based Models of Innovation Diffusion: A Critical Review.
\newblock {\em CoRR}, 1608.08517, 2016.

\end{thebibliography}

\end{document}